\documentclass{kapproc} 

\usepackage{t1enc}

\usepackage{procps} 

\usepackage[dvips]{graphicx}

\upperandlowercase

\kluwerbib 

\newcommand{\mic}{\,$\mu$m }
\newcommand{\micpa}{\,$\mu$m}          
\newcommand{\muJy}{\,$\mu$Jy }
                              
\def\araa{ARA\&A}%
\def\apj{ApJ}%
\def\apjl{ApJ}%
\def\apjs{ApJS}%
\def\mnras{MNRAS}%
\def\aap{A\&A}%
\def\physrep{Physics Reports}

\begin{document}
\def\gtapp
{\mathrel{\hbox{\raise0.3ex\hbox{$>$}\kern-0.8em\lower0.8ex\hbox{$\sim$}}}}
\def\ltapp
{\mathrel{\hbox{\raise0.3ex\hbox{$<$}\kern-0.75em\lower0.8ex\hbox{$\sim$}}}}
\def\ts{\thinspace}

\articletitle{Evolution of the IR
energy  density and SFH  up to 
$z$$\sim$1:  first results from MIPS}


\author{Emeric Le Floc'h$^1$, C.\,Papovich$^1$, H.\,Dole$^2$, E.\,Egami$^1$, P.\,P\a'erez-Gonz\a'alez$^1$, G.\,Rieke$^1$, M.\,Rieke$^1$,  E.\,Bell$^3$ \& the $Spitzer$/MIPS GTO team$^1$}
\affil{$^1$ Steward Observatory, University of Arizona, 933, N.Cherry Avenue, Tucson 85721, AZ, USA \\
$^2$ Institut d'Astrophysique
Spatiale, Universit\'e Paris Sud, F-91405 Orsay Cedex, France \\
$^3$ Max-Planck-Institut f\"{u}r Astronomie, K\"{o}nigstuhl 17, D-69117
Heidelberg, Germany}
\email{(elefloch@as.arizona.edu)}

\begin{abstract}
Using deep observations of the Chandra Deep Field South obtained with
MIPS at 24\micpa, we present our preliminary estimates on the
evolution of the infrared (IR) luminosity density of the Universe from
$z$=0 to $z$$\sim$1. We find that a pure density evolution
of the IR luminosity function is clearly excluded by the data. The characteristic
luminosity L$_{\rm IR}$* evolves at least by (1+z)$^{3.5}$ with
lookback time, but our monochromatic approach does not allow us to
break the degeneracy between a pure evolution in luminosity or an
evolution in both density and luminosity. Our results imply that 
 IR luminous systems (L$_{\rm
IR}$\,$\geq$\,10$^{11}$\,L$_{\odot}$) become the dominant population
contributing
to the comoving IR energy density beyond $z$\,$\sim$\,0.5-0.6.
The uncertainties affecting our measurements are largely dominated
by the poor constraints on the  spectral energy distributions
that are used to translate the observed 24\mic flux into 
 luminosities.
\end{abstract}


\section*{Introduction}
Deep infrared and submillimeter observations 
performed with $ISO$ and SCUBA in the late 90s revealed
a very strong evolution of IR luminous systems with lookback 
time (e.g., Smail et al. 1997, Elbaz et al. 1999).
Since these surveys were only sensitive to the brightest IR
sources,
the quantification of this evolution has yet remained
strongly debated so far. In particular, the
relative importance of
such IR-bright objects compared 
to less luminous starbursts, and their respective
contributions to the total comoving 
 energy density at high redshift
is still unclear.
$Spitzer$, the new infrared facility of NASA, 
is now providing a unique opportunity to address this issue in more detail.

\section{Infrared luminosities of MIPS sources up to $z$\,$\sim$\,1}
We performed deep observations of the Chandra Deep Field South with
MIPS at 24\mic down to a sensitivity limit of 80\muJy (5$\sigma$
detection). Cross-correlating our data with source catalogs from 
various optical surveys (i.e., VIMOS VLT Deep Survey: Le F\`evre et
al. 2004; GOODS: Vanzella et al. 2004; COMBO-17: Wolf et al. 2004;
Chandra source follow-up: Szokoly et al. 2004), we derived the
redshift of 2635 objects detected at 24\mic and mostly located at
$z$\,$\leq$1.2. We believe that this redshift identification is nearly
complete up to $z$\,$\sim$\,1.

Using  various libraries of IR luminosity-dependent spectral energy
distributions (SEDs, e.g., Dale et al. 2001, Chary
\& Elbaz 2001), we derived the total IR
luminosities of the MIPS sources from their observed 24\mic
flux. Given the poor constraint on the true SEDs characterizing these
sources at high redshift, the typical uncertainty affecting these
estimates could reach a factor of 2 to 3. More importantly, influence
of dust temperature could also add a small systematic bias
(Chapman et al. 2003) {\it that we have
not quantified so far}. We will explore this effect more thoroughly in
a forthcoming work.

Below $z$\,$\sim$\,0.5, we find that the MIPS sources are rather
modest emitters at infrared wavelengths (median L$_{\rm
IR}$=10$^{10}$\,L$_{\odot}$). At higher redshifts
(0.5\,$\ltapp$\,$z$\,$\ltapp$\,1), luminous infrared galaxies
(LIRGs, 10$^{11}$\,L$_{\odot}$\,$\leq$\,L$_{\rm
IR}$\,$\leq$10$^{12}$\,L$_{\odot}$) become  the dominant
population among the sources detected at 24\micpa, while
we also detect a significant number of more modest starbursts and
spirals. The most extreme
sources such as the ultra-luminous IR galaxies (ULIRGs, L$_{\rm
IR}$\,$\geq$\,10$^{12}$\,L$_{\odot}$) still remain quite rare up to
$z$\,$\sim$\,1.2.

\section{Evolution of the infrared luminosity function}
Based on this sample of 24\micpa-selected sources, and using the
libraries of IR SEDs for computing the $k$-corrections, we derived
monochromatic luminosity functions (LFs) at different IR wavelengths and in
various redshift bins up to $z$\,$\sim$\,1. Figure\,1 shows these
luminosity functions plotted at 60\mic and compared to the local
60\mic IRAS LF from Takeuchi et al. (2003). A strong evolution
is clearly noticeable. We find that it can not be described by
a uniform increase of the density of the  local IR galaxy population
 (``pure density'' evolution), but
requires a shift of the characteristic luminosity L* by a factor
of (1+$z$)$^{4.0\pm0.5}$. Note that the well-known degeneracy between
a pure evolution in luminosity or an evolution combining an increase
in both density and luminosity can not be broken at this stage.

\begin{figure}[ht]
\includegraphics[width=12.0cm]{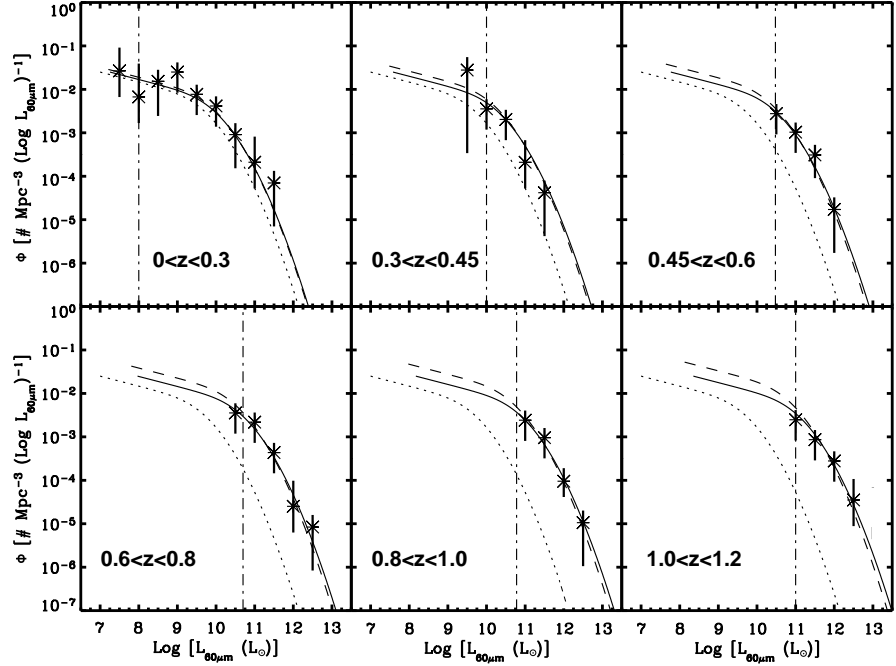}
\vskip.0in
\caption{ Luminosity functions estimated at 60\mic
with the $1/V_{\rm max}$ formalism in different redshift bins between
$z$\,=\,0 and $z$\,=\,1.2 ($\ast$~symbols). The 3$\sigma$
uncertainties are indicated with vertical solid lines.
  Data points can be fitted by a pure luminosity
evolution ($L_{\star} \propto (1+z)^{4.2}$, solid curve) or a
combination of luminosity and density evolution ($L_{\star} \propto
(1+z)^{3.5}, \Phi_{\star} \propto (1+z)^{1.0}$, dashed curve) of the
local 60\mic luminosity function (dotted curve).
 Vertical dashed-dotted lines correspond
to the 80\% completeness limit in each redshift bin.}
\end{figure}

\section{Star formation history up to $z$\,$\sim$\,1}
Star-forming galaxies are believed to be the major component
of the 24\mic source population (Silva et al. 2004).
Assuming the calibration between the star formation rate of galaxies
and their infrared luminosities (Kennicutt 1998), we converted
the evolution of the luminosity functions into an history
of the star formation up to $z$\,$\sim$\,1. Results are shown on 
Figure\,2 where the relative contributions of normal starbursts, LIRGs
and ULIRGs are also illustrated. We find that IR luminous systems (L$_{\rm
IR}$\,$\geq$\,10$^{11}$\,L$_{\odot}$) become the dominant population
 contributing
to the comoving IR energy density beyond $z$\,$\sim$\,0.5-0.6 and
 represent $\sim$70\% of the star-forming activity at $z$\,$\sim$\,1.
These preliminary results will be further developed by Le Floc'h
et al. (2005, in prep.).

\begin{figure}[ht]
\includegraphics[width=12.0cm]{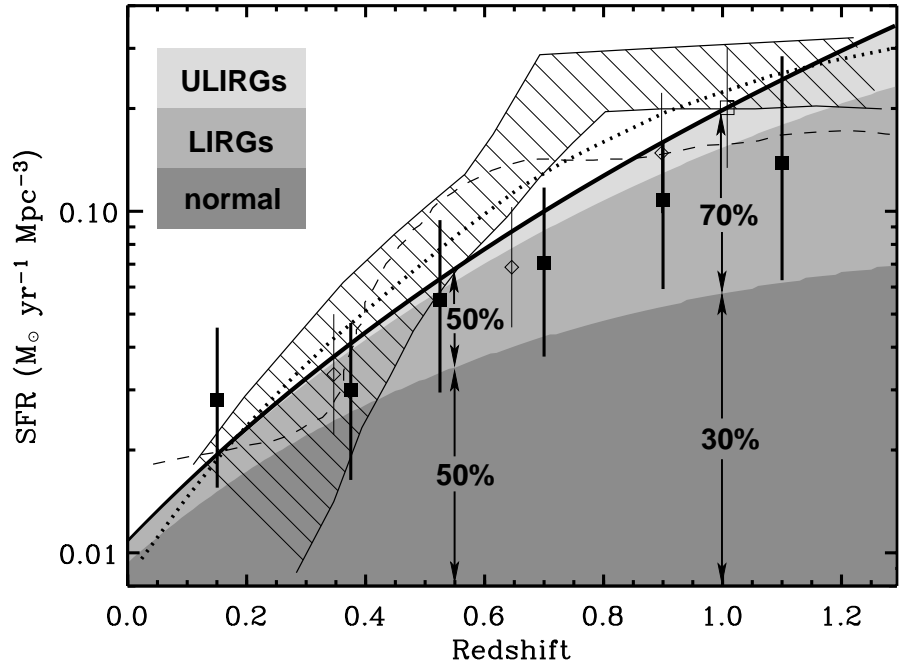}
\vskip -.0in
\caption{The star formation history up to $z$\,$\sim$\,1.2 as seen by
MIPS 24\mic (thick solid line), decomposed into the contributions of
normal galaxies (L$_{\rm IR}$\,$\leq$\,10$^{11}$\,L$_{\odot}$), LIRGs
and ULIRGs (shaded regions). Uncertainties on these estimates are still
significant  and they
are not reported for clarity. Filled squares correspond to the
contribution of $detected$ MIPS sources, while empty diamonds and the
empty square are from Flores et al. 1999 and Thompson et al. 2001,
respectively. Model predictions from Chary \& Elbaz (2001,
cross-hatched region), Blain et al. (2002, dotted line) and Lagache et
al. (2004, dashed line) are also plotted.}
\end{figure}

\end{document}